\documentclass[seceq]{ptptex}

\usepackage{graphicx}
\usepackage{wrapft}
\usepackage{verbatim}

\newcommand{\nsum}[1]{\langle\,{#1}\,\rangle}

\title{Low-dimensional chaos in populations of strongly-coupled noisy maps
}

\author{Silvia \textsc{De Monte}$^{1}$, Francesco \textsc{d'Ovidio}$^{2}$,
Erik \textsc{Mosekilde}$^3$, Hugues
 \textsc{Chat{\'e}}$^4$
}

\inst{
$^1$ CNRS-UMR 7625, Ecole Normale Sup\'erieure, 75230 Paris, France\\
$^2$ CNRS-UMR 8539, Laboratoire de M\'et\'eorologie Dynamique, Ecole Normale Sup\'erieure, 75231 Paris, France\\
$^3$ Department of Physics, 
The~Technical~University~of~Denmark, DK 2800 Lyngby, Denmark\\
$^4$ CEA -- Service de Physique de l'Etat Condens\'e, 
Centre d'Etudes de Saclay, 91191 Gif-sur-Yvette, France
}

\abst{
We characterize the macroscopic attractor of infinite populations of noisy maps
subjected to global and strong coupling by using an expansion in order
parameters.
We show that for any noise amplitude there exists a large region of strong
coupling where the macroscopic dynamics exhibits low-dimensional
chaos embedded in a hierarchically-organized, folded, infinite-dimensional
set. Both this structure and the dynamics occuring on it are 
well-captured by our expansion. In particular, even low-degree approximations
allow to calculate efficiently the first macroscopic Lyapunov exponents of the
full system.}

\begin{document}

\maketitle

\section{Introduction}

It has now been some fifteen years since the discovery of
collective behavior emerging out of infinite populations of incoherent
nonlinear/chaotic units.
In many instances, collective cycles have been found and studied, especially in
locally-coupled, spatially-extended systems. Although originally discussed 
rather early \cite{bohr87}, \ robust nontrivial collective {\it chaos} has
been documented only in the case of globally-coupled
populations,\cite{kaneko90b,matthews91,nakagawa93,nakagawa94,nakagawa95,chawanya98} \ 
and the nature and dimensionality of this chaos is still a matter of 
debate\cite{shibata98,cencini99}. Indeed, many intermediate
situations have 
been uncovered between the infinite-dimensional chaos of weakly-coupled units
and the low-dimensional chaos of the fully synchronized regimes occurring in
the strong coupling limit.

The effect of microscopic noise on such systems is a rather new topic of 
investigation.\cite{shibata99a,teramae01}
Concerning collective chaos, Shibata et al.\cite{shibata99a} showed that
its dimensionality can be drastically reduced by
microscopic noise added to populations of weakly-coupled chaotic maps.
Here we approach the same issue, starting from the fully-synchronized, 
strong-coupling limit. We have shown recently\cite{demonte04} that
this trivial collective chaos can be unfolded by the action of noise, 
allowing extra degrees of freedom to perturb the macroscopic dynamics. 
Starting from the fully-synchronized noiseless regime,
we study in some detail the apparently low-dimensional chaotic regimes
appearing when the noise intensity is increased.

This paper follows a series of previous works in which, in particular, 
we have introduced a systematic \emph{order parameter expansion} which
approximates with increasing accuracy the macroscopic dynamics observed at
strong coupling values by low-dimensional effective maps for a small set of
macroscopic observables\cite{demonte02,demonte03,demonte05}. 
Here we use this expansion scheme to tackle the issue of the actual
dimensionality of the collective chaos observed.

In the following section, we present numerical simulations illustrating
the emergence of deterministic collective behavior in the infinite-size limit
and briefly review the
order parameter expansion by which we can approximate the macroscopic
dynamics through a set of hierarchically-organized low-dimensional systems.
We then explore some consequences of such a representation in the case of
maximal coupling. Notably, we show how different microscopic
features, such as the shape of the noise distribution or the nonlinearity of
the individual dynamics, translate into the collective bifurcation diagram.

In the remaining of the paper, we address the question of the nature 
and extent of
the information about the collective behavior that we can extract 
from our low-dimensional approximations.

In Section~3, we show that the macroscopic attractor
has a folded structure that is organized with the same hierarchy as
our order parameters: the attractor is an infinitely-folded object whose
folds on a smaller scale are captured by order parameters
of increasing degree.

In Section~4, we show that the hierarchy of the reduced systems also
reflects the order in which the macroscopic degrees of freedom emerge and
determine the collective {\it dynamics}. 
The reduced systems reproduce the dominant exponents of the Lyapunov spectrum
for the population, as computed from the so-called ``nonlinear 
Perron-Frobenius operator''.\cite{kaneko92,pikovsky94a,kaneko95,morita98} 
In particular, we show that, in the region of interest,
collective chaos is characterized by only one
positive Lyapunov exponent, while the other macroscopic degrees of freedom
play the role of fast modes with negative associated Lyapunov exponents. 
Nevertheless, the
dimension of the attractor, as measured by the Lyapunov dimension, can
increase when the magnitude of the negative eigenvalues reduces. This
typically happens when the coupling is weakened.

In the last section, we discuss the perspectives of our work and in
particular the region of intermediate couplings and weak noise, where the
collective dynamics is high dimensional also in the noiseless case and the
order parameter expansion to low degree diverges.

\section{Macroscopic dynamics}

\subsection{Order parameter expansion}

We address here the population of globally-coupled noisy maps: 
\begin{equation} \label{eq:npop}
x_j\mapsto (1-K)\,f(x_j)+K\,\nsum{f(x)}+ \xi_j(t) \quad j=1,\dots N,
\end{equation}
where $f: \mathbb R \rightarrow \mathbb R$ defines the
individual (local) 
dynamics, $K\in [0,1]$ quantifies the coupling of every population element to
the average iterate $\nsum{f(x)}$ and $\xi_j(t)$ is an independent noise term,
drawn according to a given distribution of finite moments $m_q$.
The collective dynamics is addressed, 
without loss of generality, in terms of the evolution of the 
simplest macroscopic variable, i.e. the
average population state or mean field:
\begin{equation}\label{eq:mf}
X=\nsum{x}=\frac 1 N \sum_{j=1}^N x_j.
\end{equation}

Numerous studies of the noiseless case have revealed that the mean field of
such a population can display a wealth of dynamical regimes, ranging from
one-dimensional chaos for strong coupling, when full synchronization is
attained, to higher dimensional chaos in the clustering region (the
dimensionality depending on the number of clusters).
\cite{kaneko90a,kaneko92,chate91,maritan94,pikovsky94a,pikovsky94b,nakagawa98,chawanya98,manrubia00,shimada02,popovych01b,popovych02}

The addition of noise to this system ---as well as parameter diversity---
has been often conceived as a means of
testing the genericity of the aforementioned results and the structural
stability of the collective dynamics.\cite{pikovsky94a} \ 
Another approach consists in fixing the 
parameters of the individual map and systematically vary the noise
intensity. Changing the variance of the
noise distribution has revealed that the stochastic terms interact with the
nonlinearities of the dynamical system. For large populations, 
noise affects the high dimensional collective dynamics at low coupling by
reducing the dimensionality of the turbulent motion.\cite{shibata99a}

In spite of the trivial simplicity of the macroscopic dynamics in the fully
synchronous regime, the modifications of the collective behavior induced by
noise are far from straightforward. For instance, as shown by 
Kuramoto and Teramae, even weak noise can cause the
population distribution moments to scale anomalously with respect to the noise
distribution moments over a range of scales.\cite{teramae01,demonte05} 

In recent work, we have focused on how the addition of noise to a
synchronous regime modifies the collective dynamics.\cite{demonte04,demonte05}
\ Somewhat
counterintuitively, although noise blurs the trajectory of any population
element, it affects the mean field of infinite populations in a deterministic
way. This is because, in the limit of infinite population size, the 
macroscopic dynamics is the result of the averaging of
infinitely-many independent stochastic processes.
More surprisingly, the macroscopic attractor still appears
low-dimensional even away from the full-synchronization regime.

\begin{figure}[h]
\centerline{
\includegraphics[width=.8\textwidth]{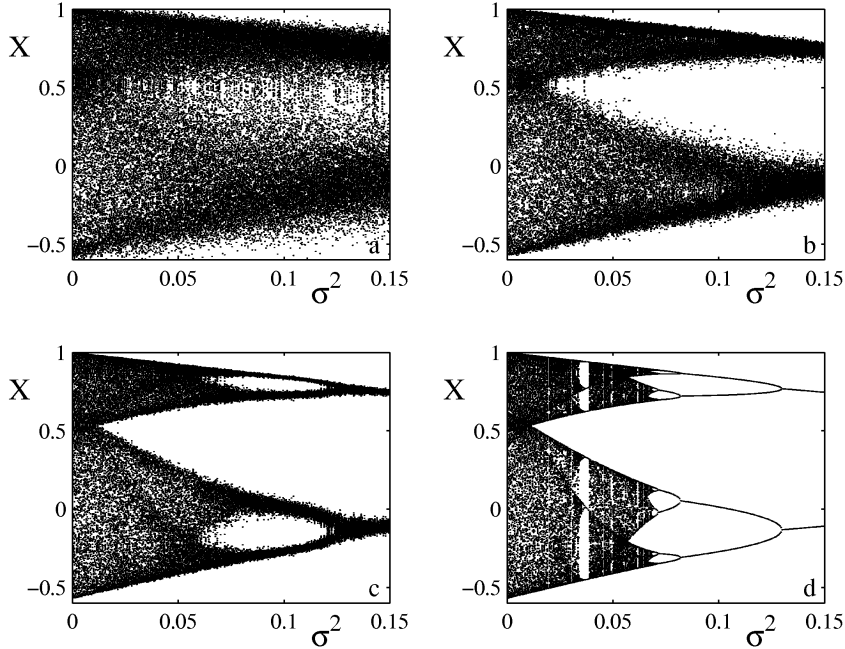}}
\caption{Bifurcation diagrams of the mean field Eq.\ (\ref{eq:mf}) for
   $K=1$ and populations of increasing size $N$: a) $N=10^2$ b)  $N=10^3$
  c) $N=10^4$ d) $N=10^6$ and reduced system. The noise intensity $\sigma^2$
  plays the role of control parameter, while the parameters of the local
  dynamics Eq.\ (\ref{eq:logistic}) are kept fixed. 
  \label{fig:sizedep}} 
\end{figure}

As an illustrative example, we consider now a population governed by
Eq.\ (\ref{eq:npop}), where
\begin{equation}\label{eq:logistic}
f(x)=1-a\,x^2
\end{equation}
is a chaotic logistic map ($a=1.57$, one-band chaos) 
and the noise distribution is taken uniform. Varying the variance of the noise
distribution $\sigma^2$ as a control parameter, a deterministic 
bifurcation diagram emerges when the population size $N$ is increased
(Fig.\ \ref{fig:sizedep}).

We recently introduced an order parameter expansion, which allows to
approximate, in the strong-coupling region, the deterministic collective
dynamics by means of a low-dimensional map.\cite{demonte04} \ 
This macroscopic effective dynamical system is obtained by writing the
evolution equation for the mean field $X$ as coupled to other 
macroscopic variables, or order parameters:
\begin{equation}\label{eq:ordpar}
\Omega_q:=\nsum{\epsilon^q}\hspace{20mm}q\in\mathbb N,
\end{equation}
where $\epsilon_j=x_j-X$ is the deviation of a population element from the
mean field.
In this case, Eq.\ (\ref{eq:ordpar}) identifies the moments of the population
distribution at a fixed time instant. 
The infinite-dimensional map defining the evolution of all such order
parameters is formulated in powers of $(1-K)$, so that a truncation to a given
degree $n$ provides a reduced system of dimension $n$ which reads: 
\begin{eqnarray}\label{eq:redpol}
\begin{cases}
X\mapsto&f(X)+\sum_{q=2}^n\:A_q(X)\Omega_q 
\\\rule[0mm]{0mm}{7mm}
\Omega_q\mapsto &
m_q+ \sum_{i=1}^q \binom{q}{i} (1-K)^i\,m_{q-i}
\,\Gamma_i (X,\Omega_2, \dots \Omega_n) \hspace{7mm} q=2,\dots n
\end{cases}
\end{eqnarray}
where $A_q$ and $\Gamma_i$ depend on the first $2P$ moments of the noise
distribution.\cite{demonte05} Of the remaining order parameters, those up to
$q=n\,P$ ($P$ being the degree of the the map $f$) have a dynamics slaved to
the the first $n$, while the others are 
constantly equal to the noise distribution moment of corresponding order.  

In the Appendix, we write the reduced systems up to the fourth degree when the
logistic map Eq.\ (\ref{eq:logistic}) defines the individual dynamics. There,
the $A_q$ and $\Gamma_i$ are explicitly derived as functions of the first
order parameters.

The population-level parameters that figure in Eqs.\ (\ref{eq:redpol}) 
are those of the individual map, defining its degree of
nonlinearity, the moments $m_q$ of the noise distribution, and the coupling
constant $K$. 
The noise variance $\sigma^2$, that we typically choose as the bifurcation
parameter, hence appears naturally as one of the parameters of the macroscopic
description.

\subsection{Interplay of map nonlinearities and noise moments}
\label{par:maxcoup}

One first consequence of representing the macroscopic dynamics by means of
Eqs.\ (\ref{eq:redpol}) is that we can disentangle the effects of the local
nonlinearities from those of the noise. 
Let us consider the case of maximal coupling $K=1$. Here, the zeroth-degree
representation for $n\rightarrow \infty$ is exact and takes
the form: 
\begin{eqnarray}\label{eq:X0}
X\mapsto 
f(X)+\sum_{q=1}^\infty\:\frac{1}{q!}\:\mathcal{D}^qf(X)\:m_q,
\end{eqnarray} 
where $\mathcal{D}^qf$ is the $q$-th derivative of the individual map.

When noise is added, the uncoupled map is perturbed by a term where the
moments of the noise distribution are weighted by the nonlinearities of the
map. Any (polynomial) map of maximal degree of nonlinearity $p$ will only
be influenced by the first $p$ moments of the noise distribution.
As a
consequence, in the case of quadratic maps such as the logistic one, the
macroscopic dynamics will be affected in the same way by any noise
distribution of given variance and is described by:
\begin{equation}\label{eq:log0}
X\mapsto 1-a\:\sigma^2-a\:X^2.
\end{equation}

This is, of course, another logistic map which can be rescaled to the uncoupled
one via a simple change of variables (see Fig.\ \ref{fig:sizedep}(d) for its 
bifurcation diagram, overlapping, up to finite-size effects,
that of the full population).

In order to explore how the macroscopic dynamics is affected by
noises of the same intensity (defined as the variance $\sigma^2$ ), but whose
distributions have otherwise different features, 
the individual map needs to have nonlinearities of order higher than two.  
Consider, for example, the quartic map defined by a perturbation of the
previously considered logistic equation: 
\begin{equation} \label{eq:quartic}
x\mapsto 1-a\,x^2+b\,x^4,
\end{equation}
where $a=1.57$ and $b=0.1$. For these parameter values, the quartic map is
chaotic. 

\begin{wrapfigure}{r}{.6\textwidth}
\centerline{
\includegraphics[width=.6\textwidth]{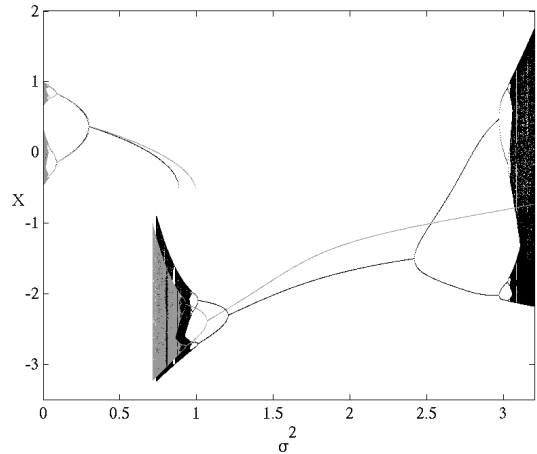}}
\caption{Bifurcation diagrams of Eq.\ (\ref{eq:quar0}) for a
  local dynamics defined by Eq.\ (\ref{eq:quartic}) with different noise
  distributions: Gaussian (grey), uniform (black). The diagrams are
  indistinguishable, up to finite-size effects, from those of the corresponding
  population. 
\label{fig:qfeig}}
\end{wrapfigure}

The reduced system Eq.\ (\ref{eq:X0}) reads in this case:
\begin{equation}\label{eq:quar0}
X\mapsto 1-a\,\sigma^2+b\,m_4-\left(a-6\,b\,\sigma^2\right)\,X^2+b\:X^4.
\end{equation}
which is another quartic map, whose bifurcation diagram is shown in
Figure\ \ref{fig:qfeig}(a) for uniform and Gaussian noise distributions.
The order parameter expansion reproduces the interaction of the single-element
nonlinearities with the noise distribution features. 

The effect on the
collective dynamics of an increase in the noise intensity allows us to
distinguish, on a purely macroscopic basis, between microscopic 
stochastic processes with differences in the moments up to the fourth one.  
The distance between the dynamics induced by the two noise distributions are
larger for larger noise intensity.
Strikingly, the collective dynamics is either chaotic or stationary
for strong noise.
Figure\ \ref{fig:qfeig}  also shows that the macroscopic dynamics is
bistable for intermediate noise values and that the
hysteresis region is slightly changed by the kind of noise
that is applied. 
Similar hysteretical phenomena have been also observed in populations of
globally coupled ODEs with independent noise,\cite{hong00} \ and find here 
a simple origin within our approach.

For any polynomial of $P$-th degree, the macroscopic dynamics at maximal
coupling will be exactly described by a polynomial of the same order, whose
parameters are the first $2P$ moments of the noise distribution. If the
uncoupled map is not polynomial, in principle all the noise distribution
moments affect the macroscopic dynamics and Eq.\ (\ref{eq:X0}) might be a
diverging series. In some cases, however, the fact that $m_q$ scales as
$\sigma^q$ allows to approximate the series by its truncation to a finite
degree in $\sigma$. As long as noise is sufficiently weak, a formal truncation
to a sufficiently high degree is able to describes in detail the macroscopic
bifurcation diagram of the full system. For instance, this
applies to population of excitable maps, where microscopic noise gives rise to
complex collective behavior.\cite{demonte04,demonte06}

\section{Fine structure of the macroscopic attractor}\label{sec:finestr}

When $K<1$, the zeroth-degree reduced system is no
longer an exact solution for the collective dynamics, but Eq.\
(\ref{eq:redpol}) provides a hierarchy of maps ordered in powers of $(1-K)$.
For not too small coupling values, the macroscopic dynamics still
looks rather low-dimensional. As we have shown already in
Refs.~\citen{demonte04} and \citen{demonte05}, this hierarchy, when truncated,
can account quantitatively for the collective dynamics.
The rest of this paper is dedicated to explore in some detail
what are the characteristics of the macroscopic dynamics that are captured by
such low-dimensional systems. In particular, we will address what
changes in the mean field attractor are induced by a reduction of the
coupling strength $K$.

\begin{wrapfigure}{r}{.52\textwidth}
\centerline{
\includegraphics[width=.5\textwidth]{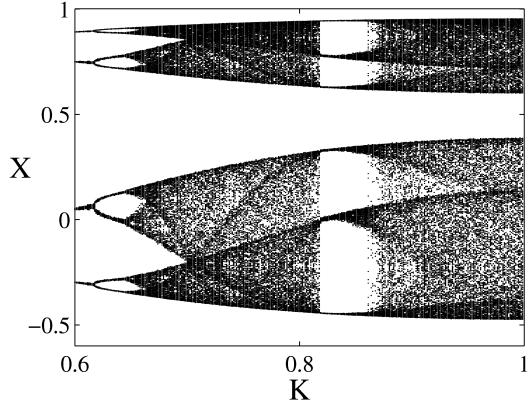}
}
\caption{Bifurcation diagram  with respect of the coupling strength
  $K$ for the mean field of a population of $N=2^{20}$ globally  
  coupled logistic maps with added microscopic uniform noise of
  variance $\sigma^2=0.03$.  \label{fig:feigs}} 
\end{wrapfigure}

Figure \ref{fig:feigs} displays a bifurcation diagram for fixed noise
intensity and varying coupling strength. As for the case, illustrated in
Fig. \ref{fig:sizedep} (d), of fixed $K$ and varying $\sigma^2$,
the collective dynamics seems to undergo macroscopic bifurcations
among low-dimensional regimes. However, the scaling in the width of
the chaotic bands and the distribution of the periodic windows suggests that
the diagram might not be straightforwardly rescaled to that of the local scalar
map. The anomalies 
become more evident when the noise intensity is weakened: for
instance, for low coupling the macroscopic dynamics enters a region
of multistability, probably associated with the presence of clusters
in the limit of zero noise.   

In this Section, we study the structure of the macroscopic
attractor and compare the simulations for the full system to those of the
reduced systems to the first four degrees. 
As a case study, we choose the population of logistic maps whose local
dynamics is defined by Eq.\ (\ref{eq:logistic}) and a uniform distribution for
the noise term. The reduced systems relative to this population are given in
the Appendix.

\begin{figure}[h]
\centerline{
\includegraphics[width=.49\textwidth]{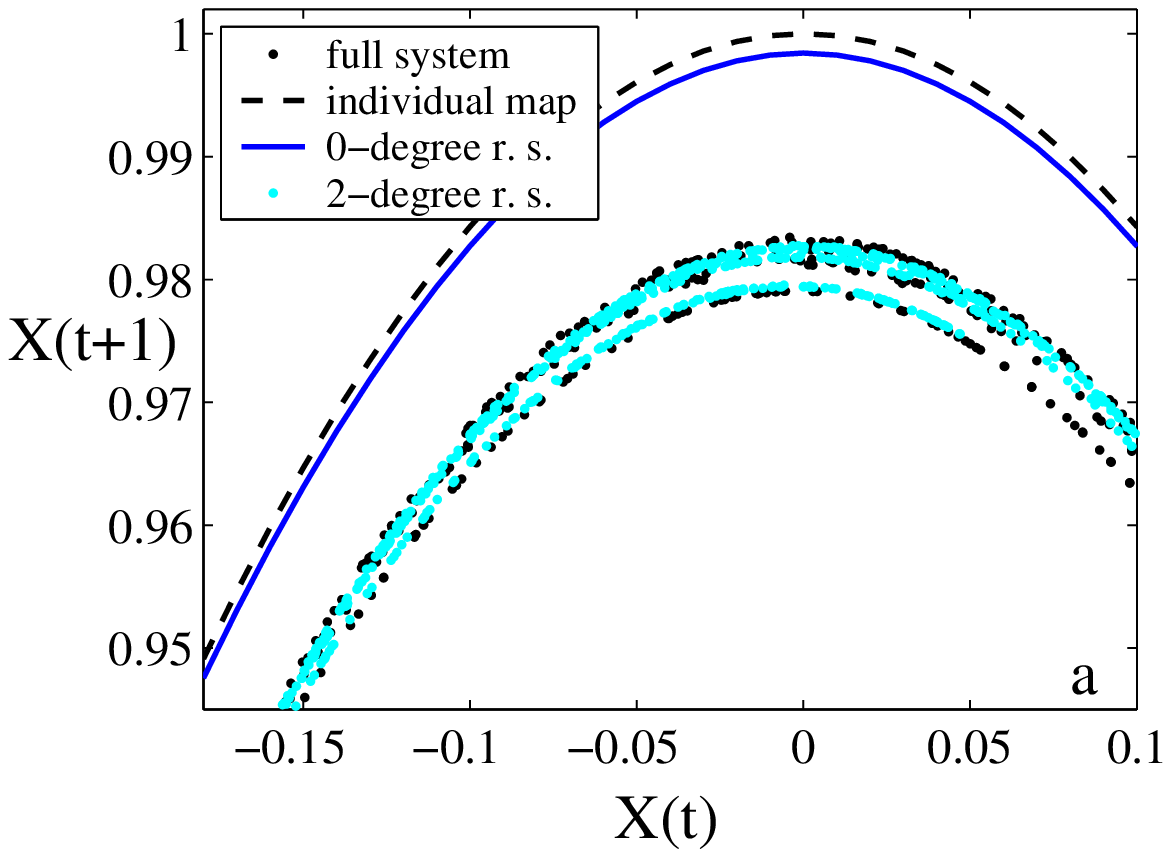}\hfill
\includegraphics[width=.48\textwidth]{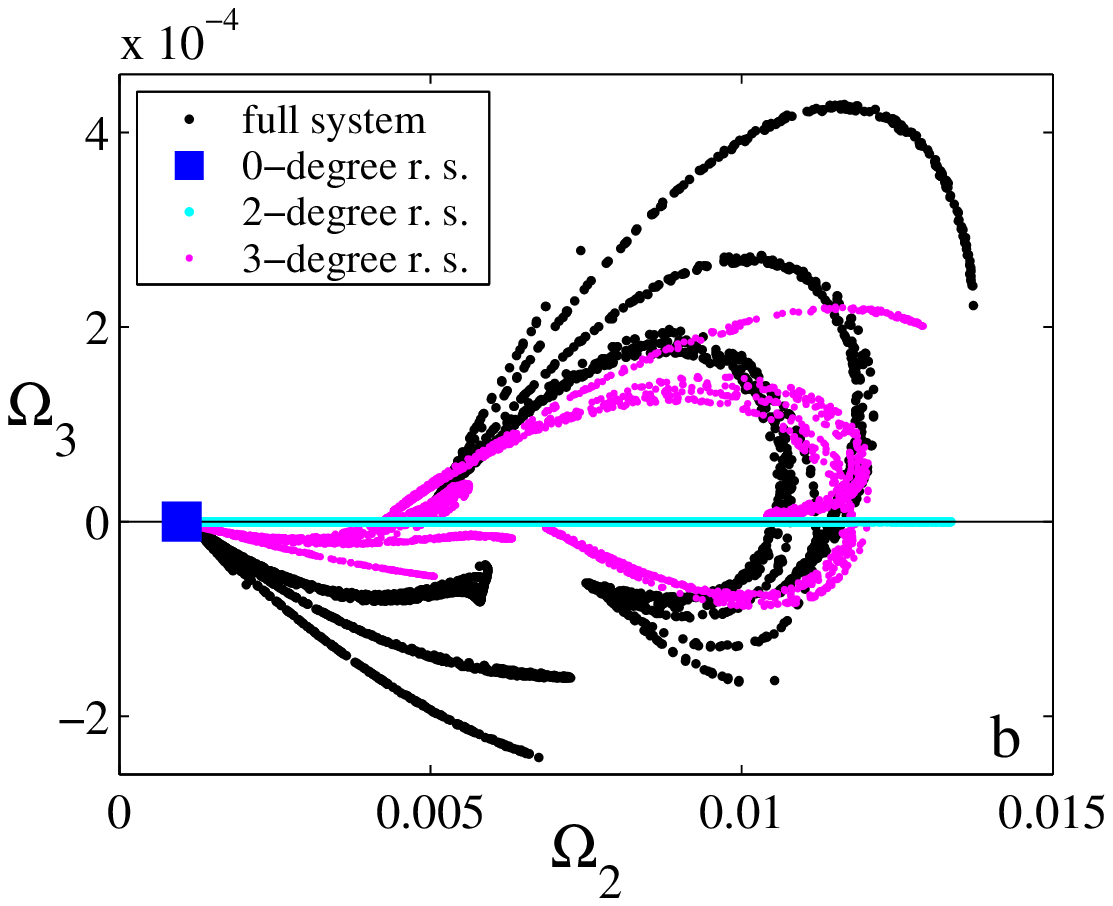}
}
\caption{a) First return map for the mean field of the population (black
  dots) and for the reduced system of second degree Eq.\ (\ref{eq:log2}) (grey
  dots), for $K=0.4$ and  
  $\sigma^2=0.001$. The dashed line is the invariant parabola Eq.\
  (\ref{eq:log0X}) where the
  single-element dynamics is embedded, the solid line is the invariant
  parabola for the zeroth-degree approximation Eq.\
  (\ref{eq:log0}).
  b) The third order parameter $\Omega_3$ versus the second $\Omega_2$ for
  the population of logistic maps (black dots) and for the third degree
  reduced system Eq.\ (\ref{eq:log3}). The continuous line
  is the invariant manifold which embeds the dynamics of Eq.\
  (\ref{eq:log2}) and of Eq.\ (\ref{eq:log0}) (square). 
\label{fig:fold}} 
\end{figure}

Let us start by considering the first return map of the mean
field. When the coupling is maximal, the mean field evolves according to the
one-dimensional map Eq.\ (\ref{eq:log0}). Its first return map hence lies onto
a one-dimensional manifold, the parabola: 
\begin{equation}\label{eq:log0X}
X(t+1)=1-a\:\sigma^2-a\:X^2(t).
\end{equation}
Reducing the coupling, this parabola folds (Fig.\ \ref{fig:fold}(a)), which
indicates that the macroscopic attractor is no longer strictly
one-dimensional.   
Eq.\ (\ref{eq:log0}) is not able to account for the modification of
the attractor, but the folding is captured with great
accuracy by the reduced system of second degree Eq.\ (\ref{eq:log2}).

In spite of the fact that the reduced system of second degree reproduces
qualitatively and quantitatively the macroscopic dynamics of the population,
this approximation is not able to track modifications of the mean field
attractor related to the asymmetry of the population distribution around its
average. Such asymmetry appears as the coupling is less than maximal, and is
best revealed by looking at the third order parameter $\Omega_3$ as a function
of the second order parameter $\Omega_2$ (Fig.\ \ref{fig:fold} (b)). 
In the zeroth and second-order approximations, the dynamics
of $\Omega_2$ is restricted to the line $\Omega_3=0$, and in the first case it
reduces to the trivial fixed point $(\sigma^2,m_3)$ of Eq.\ (\ref{eq:log0}). 
The reduced system of third degree is a three-dimensional map, reported in
Appendix, for the mean field, the second and the third order
parameter. Figure\ \ref{fig:fold} (b) shows  that, contrary to the lower
degree truncations, it gives rise to the same folding observed for the
population third moment.   

\begin{figure}[h]
\centerline{
\includegraphics[width=.48\textwidth]{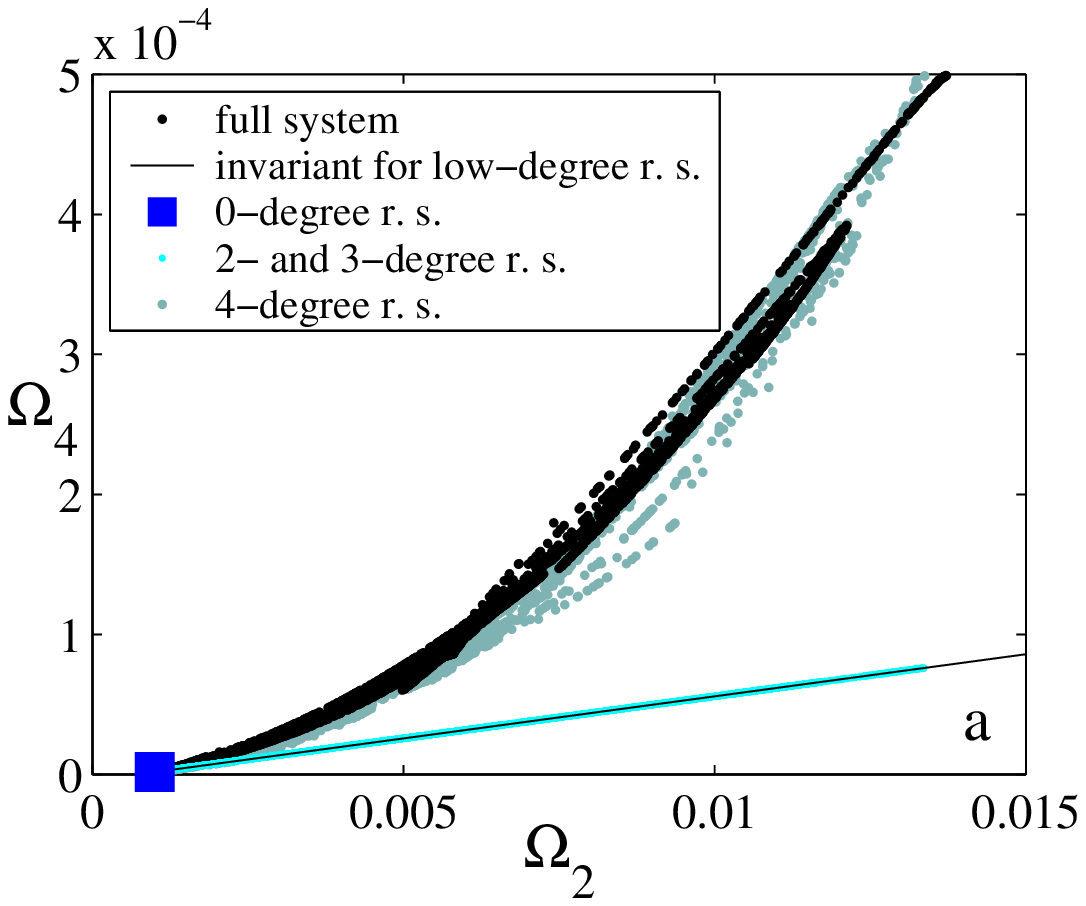}\hfill
\includegraphics[width=.49\textwidth]{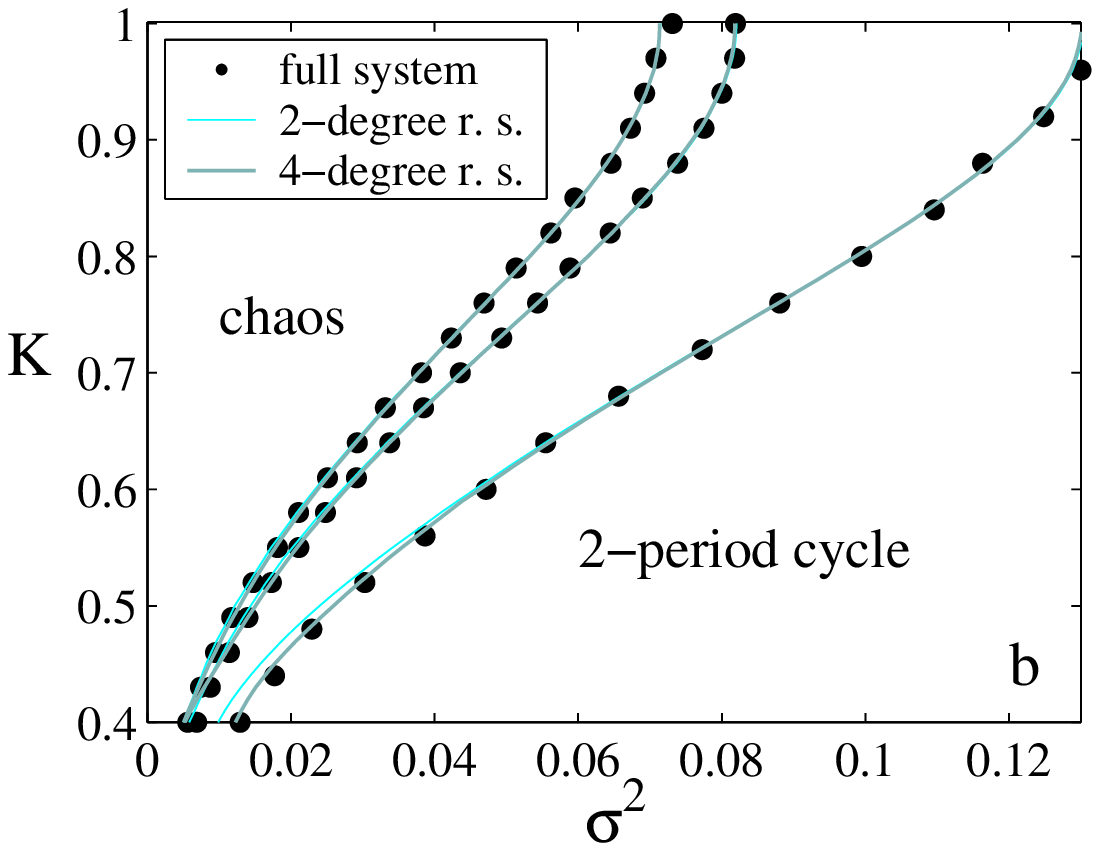}
}
\caption{
  a) The fourth order parameter $\Omega_4$ versus the second $\Omega_2$ for
  the population of logistic maps (back dots) and for the
  reduced system of fourth degree Eq.\ (\ref{eq:log4}). The
  continuous line 
  is the invariant manifold which embeds the dynamics of the second and
  third degree Eqs.\ (\ref{eq:log2}) and\ (\ref{eq:log3})
  and of the reduced system of zeroth degree Eq.\
  (\ref{eq:log0}) (square). 
  b) First period-doubling bifurcation lines for the full (dots), for
  reduced system of second degree Eq.\ and
  fourth   degree Eq.\ (\ref{eq:log4}). The zeroth degree
  approximation would give vertical lines corresponding to the bifurcation
  values for $K=1$.
\label{fig:opfold}} 
\end{figure}

Finally, we project the macroscopic attractor onto the plane
$(\Omega_2,\Omega_4)$, as shown in Fig.\ \ref{fig:opfold}(a). 
In the reduced system at zeroth degree, the two order parameters coincide with
the noise distribution moments and thus identify with the point
$(\sigma^2,m_4)$. 
In the reduced systems of both second and third degree, instead, the dynamics
of $\Omega_4$ is embedded in the invariant line defined by the conservation
relation Eq.\ (\ref{eq:log2_o4}). 
In the reduced system of fourth degree in $(1-K)$, the fourth order parameter
is an independent variable. Figure\ \ref{fig:opfold}(a) shows that its
dynamics is very close to that of the population and displays a similar folded
structure.  

As the degree of the truncation is increased, one observes a convergence
in the dynamics of the lower-order variables. Thus,
the reduced system of fourth degree does not visibly improve
the agreement of the lowest degree order parameters with the macroscopic
variables of the population. As expected, the fourth degree approximation
is significantly better only at low coupling values, 
close to the clustering region. This is clear at weak noise values,
when this reduced system provides a better approximation
for the macroscopic bifurcations diagram than the lower degree ones (Fig.\
(\ref{fig:opfold}(b)).  

The distance between the full system attractor and that of 
the reduced system of lowest degree can be computed as the mean square 
distance of $X$ from the parabola 
defined by Eq.\ (\ref{eq:log0X})). It scales as $(1-K)^2$ for fixed
noise intensity in a large region of the
parameter space.\cite{demonte04} \ This confirms that the approximation to
second degree indeed captures the most important deviation of the population
dynamics from the reduced system of zeroth degree. 
Similarly, the distance of the fourth order parameter from the line Eq.\
(\ref{eq:log2_o4}), valid for the approximations to second and third degree,
scales as $(1-K)^4$ when the coupling strength is varied.

We conclude that our order parameter expansion describes well the fine
structure of the macroscopic attractor. In 
particular, the fact that we can unveil finer and finer folds by adding higher
order parameters suggests that the hierarchy of the folds is well captured by
the hierarchy of the reduced system. 
We now turn our attention to the {\it dynamics}
occurring on this complex object, trying to quantify the dimensionality of the
macroscopic chaos observed and to assess to what extent it is reproduced by
our approximations.

\section{Lyapunov spectrum of the macroscopic dynamics}

In this section, we show that the Lyapunov exponents
of the reduced system correspond exactly 
to the largest macroscopic Lyapunov exponents of the full system. 

These largest macroscopic Lyapunov exponents are most easily computed from
the dynamics of the full system expressed in terms of the evolution
of the probability distribution function (pdf) $\rho_t(x)$ of all
elements at time $t$. This pdf is governed by the so-called
nonlinear Perron-Frobenius operator:\cite{pikovsky94a,shibata99a,shibata99b} \    
\begin{equation}
\rho_{t+1}(x)=\int{G\left\{F_t(y)-x\right\}\,\rho_t(y)\,dy}
\end{equation}
where $G\{z\}$ is the noise distribution and $F_t$ the iterate of the map:
\begin{equation*}
F_t(y)=\left(1-K\right)\,f(y)+K\int \rho_t(z)\,dz. 
\end{equation*}
Numerically, this is implemented easily by choosing a
sufficiently fine binning of the support of the pdf, which is then represented
by a vector. The Lyapunov exponents are computed from this discretized system
with the method of Benettin et al.,\cite{benettin80} \  
following the evolution of a step perturbation of the pdf, according to the
formula: 
\begin{equation}\label{eq:lyapexp}
\lambda_i=\frac 1 T \sum_{t=1}^T \log (\Delta_i^t),
\end{equation}
where $\Delta_i^t$ is the ratio between the Euclidean norm of the
$i$-th component of a vector basis at time $t-1$ and its
corresponding component of the basis at time $t$. This last basis
is recomputed at each time step, after a sufficiently long transient,
by orthonormalizing the image of the basis of the tangent space at
time $t-1$.

We compute the Lyapunov exponents of the reduced system by evaluating
explicitly the Jacobian of the system and by computing the iterates of vectors
belonging to the tangent space. For the case of the Perron-Frobenius operator,
instead, we iterate perturbations of small size along a trajectory. The
perturbation size has to be chosen such that the dynamics remains close to the
linear regime and a sufficient numerical precision is maintained. The
perturbation size appeared not to be a critical parameter, 
except for the region close to maximal coupling $K=1$.
\begin{wrapfigure}{r}{.5\textwidth}
\centerline{
\includegraphics[width=.5\textwidth]{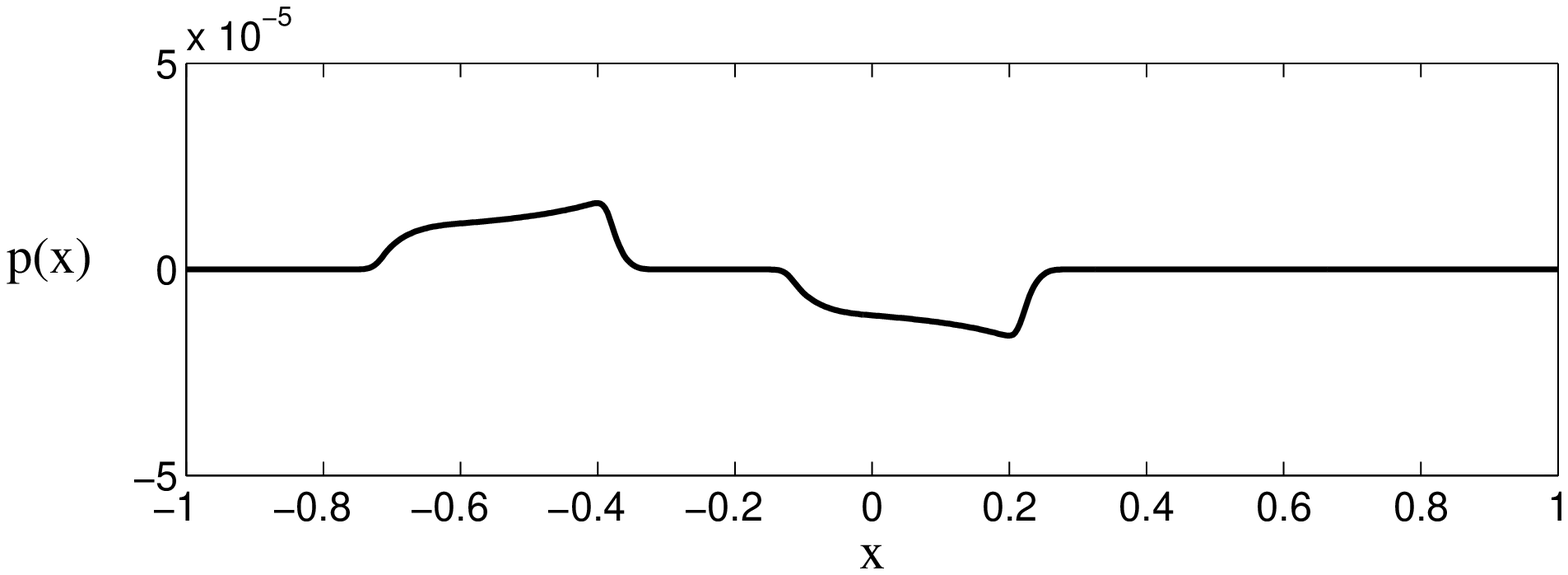}}

\centerline{
\includegraphics[width=.5\textwidth]{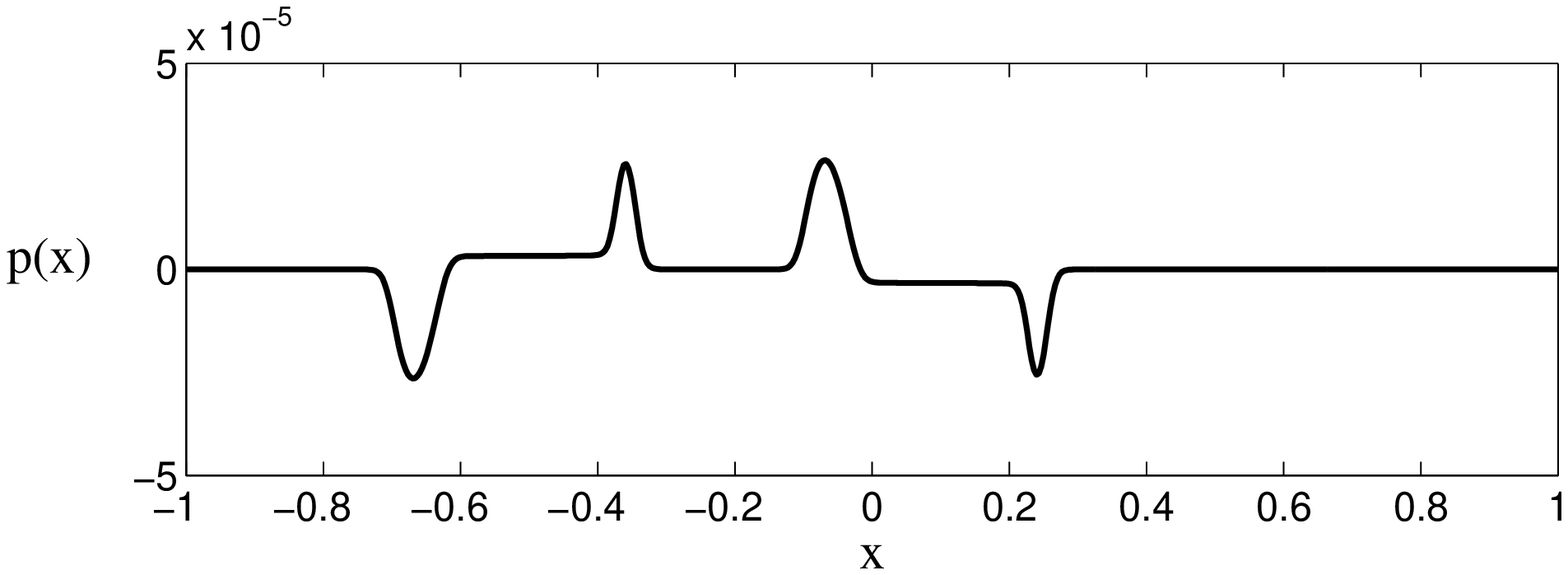}}

\centerline{
\includegraphics[width=.5\textwidth]{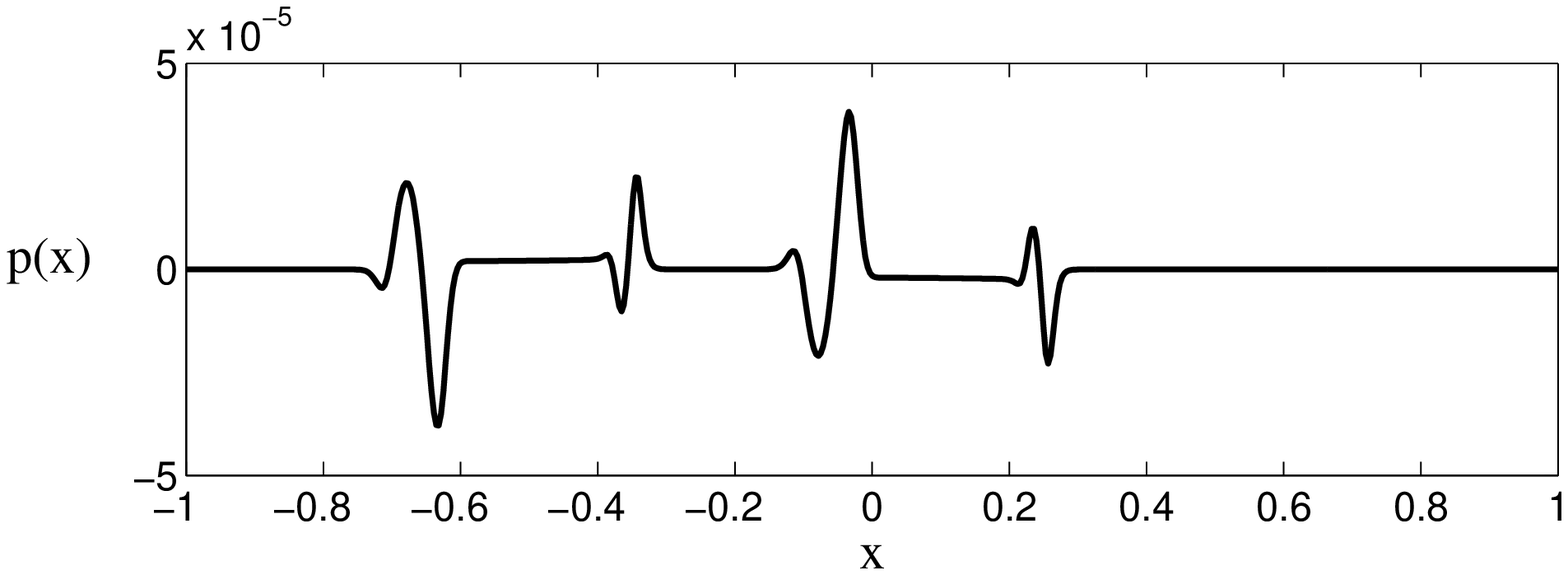}}

\centerline{
\includegraphics[width=.5\textwidth]{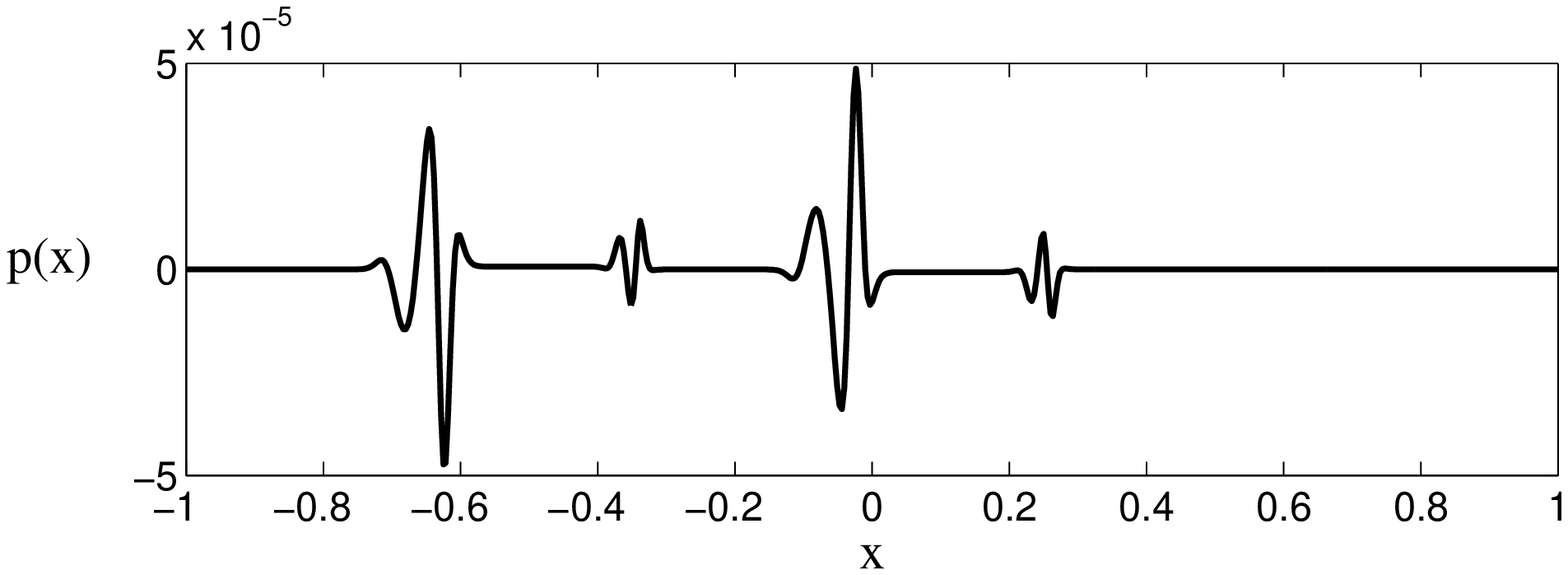}}

\caption{Eigenvectors relative to the four largest Lyapunov exponents
  for the population of noisy logistic maps ($K=0.5$ and $\sigma^2=0.03$).
  \label{fig:lyapeigenv}} 
\end{wrapfigure}
In this region there
are directions converging very fast, so that one time step may be
sufficient to greatly flatten the small-scale structures of a distribution,
leading to a decrease of numerical precision. 
This effect appears as a small deviation between the full system and the
reduced one for the higher moments when $K$ approaches $1$. In the following,
we have checked that our results do not depend on the level of discretization
of the support, insuring the extensivity of the spectrum.

In Fig.\ \ref{fig:lyapeigenv} we show the eigenvectors (in
the space of the pdfs) relative to the dominant Lyapunov exponents 
in a typical case. The increasing steepness of these objects as one goes down
the Lyapunov spectrum signals that a reliable determination of many exponents
is a computationally demanding task.

\begin{figure}[h]
\centerline{
\includegraphics[width=.48\textwidth]{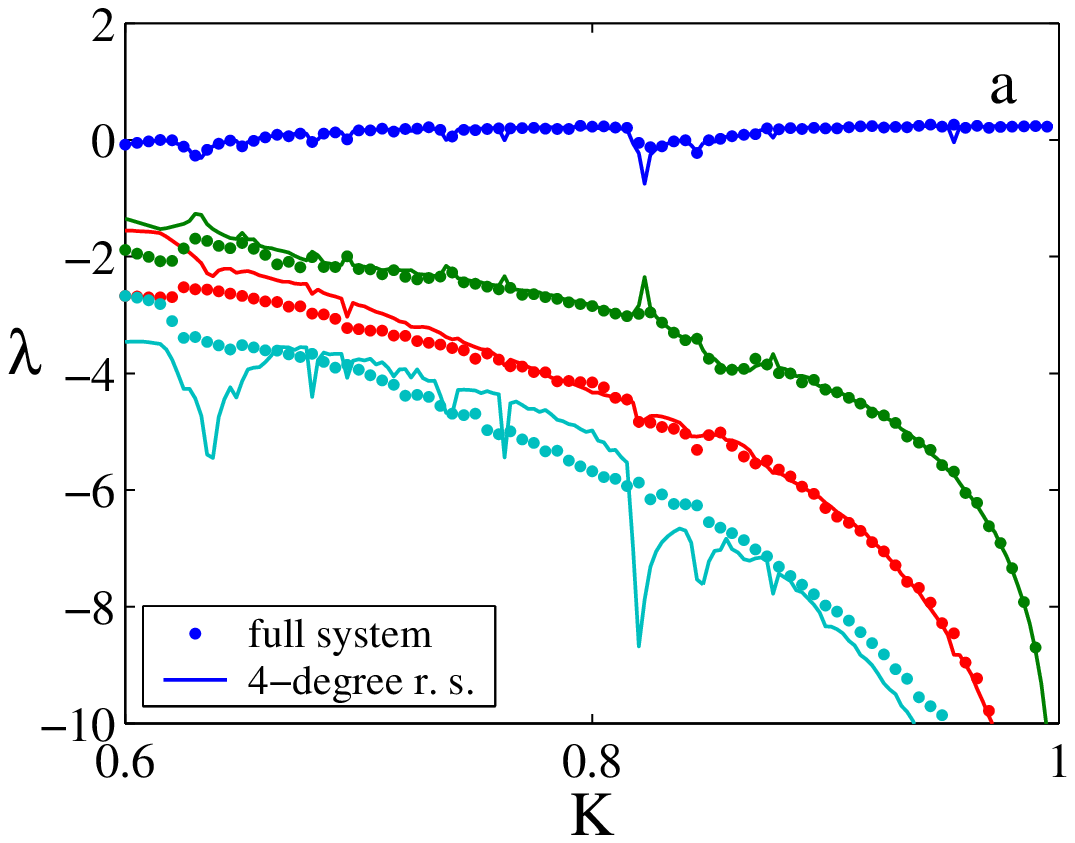}\hfill
\includegraphics[width=.48\textwidth]{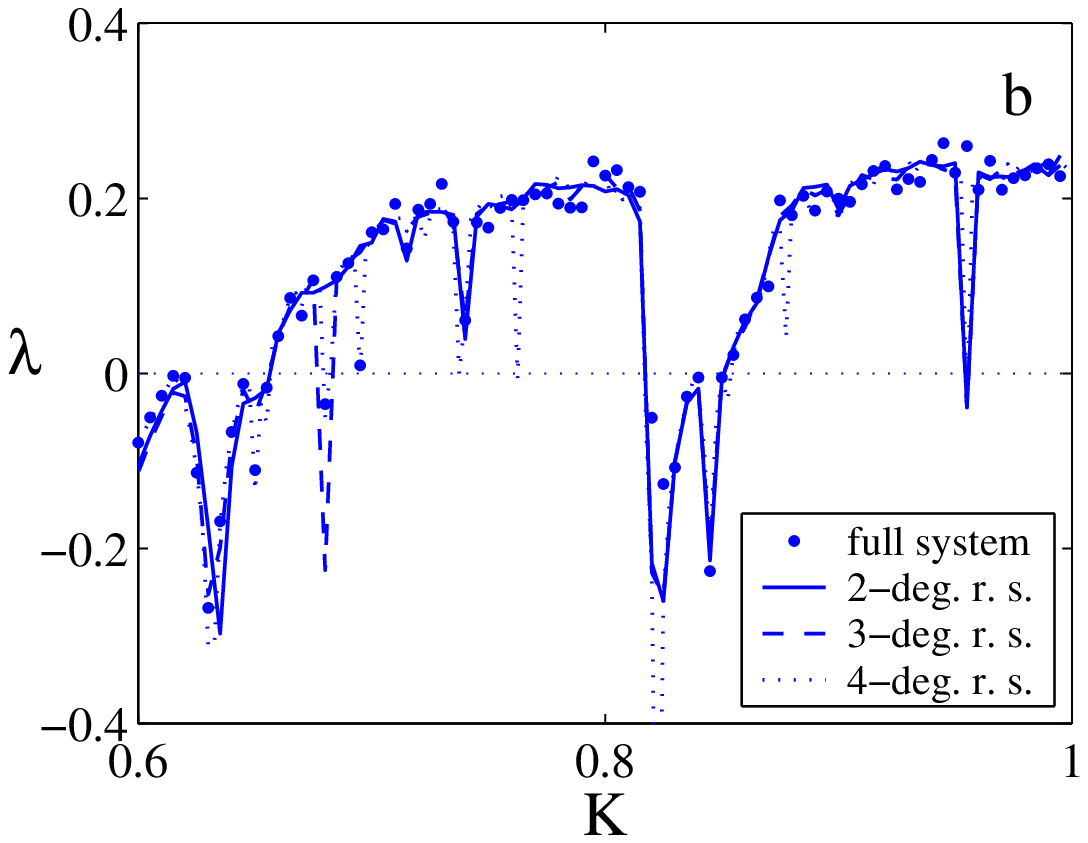}
}
\caption{a) Lyapunov exponents of the population (dots) and of the reduced
  system of fourth degree (solid lines) as a function of the coupling
  strength, for fixed noise intensity $\sigma^2=0.03$.
  b) Maximal Lyapunov exponent of the population (dots) and of the reduced
  systems of second (solid line), third (dashed line) and fourth degree
  (dotted 
  lines) as a function of the coupling strength, for fixed noise intensity
  $\sigma^2=0.03$. The fourth degree truncation provides a slightly better
  approximation for the lower values of the coupling in the displayed range.
  \label{fig:lyapcfr}} 
\end{figure}

In Fig.\ \ref{fig:lyapcfr}(a), we show the four largest Lyapunov exponents 
for the full system and for the reduced system of fourth degree as a
function of the coupling strength $K$ and for fixed noise 
intensity. Clearly, the reduced system captures the
main features of the macroscopic dynamics, i.e. the existence of only one
positive Lyapunov exponent (magnified in Fig.\ \ref{fig:lyapcfr}(b)). The
quantitative agreement with the population is extremely good for the first two
exponents. The discrepancy on the third and fourth 
exponent, evident for couplings close to the maximal one, are due to a
degradation of the numerical accuracy when some directions are too
strongly attractive. The quantitative correspondence between the
population and the reduced system, especially as far as the third and
fourth exponents are concerned, is also lost for too low values of
the coupling, where the introduced approximation is likely to be
insufficient to take every detail into account. 

\begin{wrapfigure}{r}{.48\textwidth}
\centerline{
\includegraphics[width=.48\textwidth]{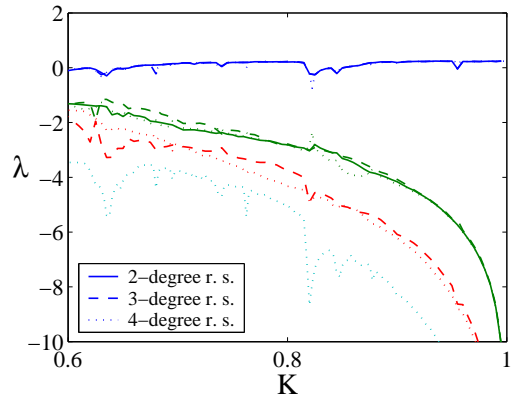}
}\caption{Lyapunov exponents of the reduced system to second (two solid lines),
third (three dashed lines) and fourth degree (four dotted lines) 
as a function of the coupling strength, for fixed noise intensity $\sigma^2=0.03$.
\label{fig:opelyapcfr}} 
\end{wrapfigure}

In the $K\to 1$ limit the maximum Lyapunov exponent tends to the value which
can be derived from the exact scalar map Eq.\ (\ref{eq:X0}), while
all the other directions are all the more attracting as $K$ is close to 1.
The truncations of degree greater than zero provide the successive exponents,
each approximation level adding one new exponent which is more negative than
those retrieved with the truncations to lower degree. This can be seen in Fig.\
\ref{fig:opelyapcfr}, which displays the Lyapunov exponents of
the reduced system at different truncation levels.

In view of the obtained quantitative and qualitative agreement with
the Lyapunov exponents of the population, we conclude that, in spite
of being just one of the 
possible projections of the population dynamics on a lower dimensional space,
the order parameter expansion indeed captures the hierarchical structure of
the macroscopic dynamics.

\subsection{Dimension of the macroscopic attractor}

The computation of the Lyapunov exponents also allows us to address the
delicate problem of how the macroscopic attractor dimension depends on the
population parameters $K$ and $\sigma$. 
The question of whether the collective behavior is truly 
low-dimensional, or instead always infinite-dimensional, has been mainly
addressed in the context of weakly-coupled noiseless 
maps.\cite{kaneko95,nakagawa98,chawanya98,nakagawa99,topaj01} \ Shibata,
Chawanya and Kaneko pointed out that, for weak coupling, it is sufficient to
add a small amount of noise to reduce drastically the dimensionality of the
collective motion, 
measured as the number of positive Lyapunov exponents of the spectrum of the
Perron-Frobenius operator.\cite{shibata99a,shibata99b}

Here we approach this question starting from maximal coupling 
(full synchronization) rather than
from zero coupling and compare the attractors of the full and reduced systems
on the basis of their Lyapunov or Kaplan-Yorke dimension:   
\begin{equation}\label{eq:kydim}
D_{L}=j+\frac{\sum_{k=1}^j\lambda_k}{|\lambda_{j+1}|}
\end{equation}
where $j$ is the smallest integer such that, if the Lyapunov exponents
$\lambda_k$ are listed in descending order,
$\sum_{k=1}^j\lambda_k\ge 0$.
The Kaplan-Yorke conjecture, which states that the fractal dimension of a
strange attractor is equal to $D_{L}$ \cite{ottbook} appears to hold well for
sufficiently 
nonsingular maps, and has been proved to be an upper estimate of the fractal
dimension for a class of maps. \cite{grassberger83} \ We thus expect to be able
to use the Lyapunov dimension as a measure of the dimension of the macroscopic
attractor. 

\begin{wrapfigure}{r}{.5\textwidth}
\centerline{
\includegraphics[width=.5\textwidth]{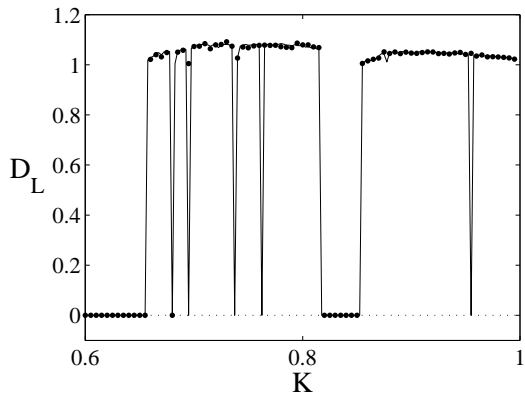}
}
\caption{Lyapunov dimension of the macroscopic attractor for the population
  (dots) and the reduced system to the fourth degree (solid line). The scan
  corresponds to the bifurcation diagram Fig.\ \ref{fig:feigs}, for
  $\sigma^2=0.03$. The values 
  of the Lyapunov dimension of the reduced systems of lower degree nearly
  overlap with those of the fourth degree, and therefore are not displayed.
  \label{fig:kydim}} 
\end{wrapfigure}

Although the addition of more and more macroscopic degrees
of freedom improves the description of the mean field attractor by revealing
smaller scale structures, as we discussed in Section\ \ref{sec:finestr}, this
does 
not necessarily lead to an increase in the dimension of the macroscopic
attractor. Indeed, in the region of interest here,
the macroscopic attractor is always characterized by one
positive Lyapunov exponent and its increase in dimension is associated to
changes of the attractiveness of the transversal modes. If the coupling is
sufficiently strong, the dimension of the attractor is less than two and is
determined only by the first one of these modes, that associated with the
second order parameter. 

In Fig.\ \ref{fig:kydim}, we show that the Lyapunov dimension of the reduced
system has the same dependence on the population parameters as
that of the population. 
In particular, this dimension tends to one for $K\to 1$, that is where
we have analytically concluded that the macroscopic attractor is strictly
one-dimensional. 
Weakening the coupling, instead, one observes an increase in the
attractor dimension, until the system exits the chaotic region through a
macroscopic bifurcation, and the
Lyapunov dimension drops to zero. Similarly, the intervals within 
the chaotic region in which the dimension of the macroscopic attractor is zero
correspond to the windows of periodic behavior that one can identify in the
bifurcation diagram shown in Fig.\ \ref{fig:feigs}.

\section{Discussion}

We have studied the macroscopic attractor of large populations of noisy
maps. We have shown that, for
sufficiently strong coupling, there exists a truncation of the order
parameter expansion which is able to reproduce the collective dynamics to any
given scale of description. 
For a fixed coupling, the macroscopic dynamics is effectively described by a
sufficiently large number $n$ of order parameters. In this approximation, some
of the following order parameters are slaved to the first ones, while all the
others can be approximated by the noise 
distribution moments within the error of a macroscopic description. 
The hierarchy of reduced systems obtained in such a manner reflects the
fractal structure of the mean field attractor, that can be described on a
finer and finer scale as the degree of approximation is increased.
In spite of its complex structure embedded into an infinite-dimensional phase, 
however, the macroscopic attractor has a finite Lyapunov dimension, which is 
the same as that of the reduced systems.

Some further questions that can be addressed by means of the reduced systems
are whether the macroscopic attractor attains its maximum Lyapunov
dimension for fixed noise intensity and how this maximum value changes with
$\sigma^2$. In particular, it would be interesting to determine whether this
maximal dimension converges to a finite value in
the limit of weak noise for any coupling strength.

From the order parameter expansion, it is clear that the moments of the noise
distribution constitute the most relevant terms in the evolution of
the order parameter of corresponding degree. In the limit of maximal coupling,
thus, the order parameter of $q$-th degree scales ``normally'', that is
proportionally to $\sigma^q$. For coupling less than maximal, these iterates
are 
perturbed by the macroscopic dynamics, so that one can identify deviations
from the normal scaling associated with the bifurcations of the mean field
attractor. 
It is in principle conceivable that such bifurcations give rise to the
``anomalous scaling'' observed for intermediate coupling strengths, close to
the point where the synchronous regime breaks down for a population of
noiseless maps. Another possibility is that for intermediate couplings the
effective dimensionality of the mean field dynamics diverges, so that no
finite truncation is able to capture the scaling of the order parameters with
the noise intensity.
In order to answer this question, we are now exploring the region of low
coupling/weak noise 
close to the turbulent regime, where the attractors for the low-dimensional
reduced systems presented in this paper become unstable.

\section*{Acknowledgments}
S. De Monte acknowledges support from the ESF programme REACTOR and the EIF
010169 Marie Curie fellowship. 

\appendix
\section{Reduced systems up to fourth degree for logistic maps}

We explicitly derive the reduced systems for the case where individual maps
have the quadratic form Eq.\ (\ref{eq:logistic}). The explicit form of $A_q$
and $\Gamma_i$ are obtained following the procedure illustrated in the
derivation of Eq.\ (\ref{eq:redpol}) \cite{demonte05}: first, we write the
equations for the first $2n$ moments and then we eliminate $n$ 'slaved'
degrees of freedom by noticing that certain linear combinations of the order
parameters are invariant. 

As we have seen in the paragraph \ref{par:maxcoup}, Eq.\
(\ref{eq:redpol}) to the zeroth degree yields Eq.\ (\ref{eq:log0}). 
 
The second degree reduces system reads:
\begin{align}\label{eq:log2}\begin{cases}
X\mapsto & 1-a\,X^2-a\:\Omega_2\\
\Omega_2\mapsto & \sigma^2+(1-K)^2\:a^2
\left[m_4-6\sigma^4+(4X^2-\Omega_2+6\:\sigma^2)\:\Omega_2\right]
\end{cases}
\end{align}
and the two conservation relations are valid:
\begin{eqnarray}
\Omega_3=&&m_3 \nonumber
\\
\Omega_4=&&m_4+6\,\sigma^2\left(\Omega_2-\sigma^2\right).\label{eq:log2_o4}
\end{eqnarray}

The third degree reduced system is the three-dimensional macroscopic map:
\begin{align}\label{eq:log3}\begin{cases}
X\mapsto & 1-a\,X^2-a\:\Omega_2\\
\Omega_2\mapsto & \sigma^2+(1-K)^2\:a^2\:\Sigma_2\\
\Omega_3\mapsto & -(1-K)^3\:a^3\:\Sigma_3 
\end{cases}
\end{align}    
where:
\begin{eqnarray*}
\Sigma_2=&& m_4-6\,\sigma^4+6\,\sigma^2\,\Omega_2-\Omega_2^2+4\,X^2\,\Omega_2-4\,X\,\Omega_3 \\
\Sigma_3=&& m_6-15\,m_4\,\sigma^2+\left(12\,m_4-72\,\sigma^4\right)\,X^2
+\left(12\,m_4+18\,\sigma^4\right)\,\Omega_2+72\,\sigma^2\,X^2\,\Omega_2\\
&&-18\,\sigma^2\,\Omega_2^2
-12\,X^2\,\Omega_2^2+2\,\Omega_2^3-12\,X\,\Omega_2\,\Omega_3+8\,X^3\,\Omega_3.
\end{eqnarray*}  

The following three order parameters are slaved to the first three macroscopic
variables and they obey the relations:
\begin{eqnarray*} 
\Omega_4=&& m_4+6\,\sigma^2\left(\Omega_2-\sigma^2\right)\label{eq:log3_o4}
\\
\Omega_5=&& 10\,\sigma^2\,\Omega_3\nonumber \\
\Omega_6=&& m_6+15\,m_4\left(\Omega_2-\sigma^2\right),\nonumber
\end{eqnarray*}  
while $\Omega_q=m_q$ holds for $q>6$. 
\bigskip

The truncation of the order parameter expansion to the fourth degree is the
four-dimensional macroscopic map:
\begin{align}\label{eq:log4}
\begin{cases}
X\mapsto & 1-a\,X^2-a\:\Omega_2\\
\Omega_2\mapsto & \sigma^2+(1-K)^2\:a^2\:\Sigma_2\\
\Omega_3\mapsto & (1-K)^3\:a^3\:\Sigma_3 \\
\Omega_4\mapsto &
m_4+6\:\sigma^2\:(1-K)^2\:a^2\Sigma_2+(1-K)^4\:a^4\:\Sigma_4
\end{cases}
\end{align}    
where:
\begin{eqnarray*}
\Sigma_2=&& \Omega_4-\Omega_2^2+4\,X^2\,\Omega_2+4\,X\,\Omega_3 \\
\Sigma_3=&& \Omega_6+6\,X\,\left(\Omega_5-2\,\Omega_2\,\Omega_3\right)
-3\,\Omega_2\,\Omega_4+8\,X^3\,\Omega_3
-12\,X^2\,\left(\Omega_2^2-\Omega^4\right)+2\Omega_2^3\\
\Sigma_4= && \Omega_8
+8\,X\,\left(3\,\Omega_2^2\,\Omega_3-2\,\Omega_2\,\Omega_5+\Omega_7\right)
+24\,X^2\left(\Omega_2^3-2\,\Omega_2\,\Omega_4+\Omega_6\right)-4\Omega_2\,\Omega_6\\
&&-32 X^3\left(\Omega_2\,\Omega_3-\Omega_5\right)
+6\Omega_2^2\,\Omega_4+16 X^4\,\Omega_4-3\Omega_2^4.
\end{eqnarray*}    

The successive order parameters depend linearly on the first ones according to
the relations:
\begin{eqnarray*} 
\Omega_5=&& 10\,\sigma^2\,\Omega_3\\
\Omega_6=&& m_6+\left(m_4-6\,\sigma^4\right)\left(\Omega_2-\sigma^2\right)
+\sigma^2\left(\Omega_4-m_4\right)\\
\Omega_7=&& 35\,m_4\,\Omega_3\\
\Omega_8=&& m_8+28\,\left(m_6-15\,m_4\,\sigma^2\right)
\left(\Omega_2-\sigma^2\right)+70\,m_4\left(\Omega_4-m_4\right)
\end{eqnarray*}  
and the remaining order parameters are equal to the noise distribution moments.


\end{document}